# Impact Analysis of COVID-19 in Bangladesh Power Sector and Recommendations based on Practical Data and Machine Learning Approach

Anis Ahmed[a,∂], Arefin Ahamed Shuvo[a,∂], Naruttam Kumar Roy[a], Neloy Prosad Bishnu[a], Ali Nasir[b,c,*]

[a]Department of Electrical and Electronic Engineering, Khulna University of Engineering & Technology, Khulna-9203, Bangladesh.

[b]Control and Instrumentation Engineering Department, King Fahd University of Petroleum and Minerals, Dhahran, Saudi Arabia.

[c]Interdisciplinary Research Center for Intelligent Manufacturing and Robotics, King Fahd University of Petroleum and Minerals, Dhahran, Saudi Arabia.

*ali.nasir@kfupm.edu.sa

[∂]**Contributed equally to this study**
**\*Corresponding author**

## Abstract

This paper investigates the impact of COVID-19 on the power sector in Bangladesh, how the country has dealt with it, and explores the path to stability. The study employs data visualisation and complex statistics to examine critical data about power systems in Bangladesh. This includes load patterns on a daily, monthly, annual, weekend, and weekday basis. Significant alterations in these patterns have been observed during our study e.g., in April and May of 2020, the power demand decreased by approximately 15.4% and 17.2%, respectively, compared to the corresponding period in 2019. We have used a Long-Short-Term Memory (LSTM) framework to predict the load profile of 2020 excluding COVID-19 effects. This model is compared with the actual load profile to determine the degree to which COVID-19 has impacted. The comparison indicates that the average power demand decreased by approximately 19.5% in April 2020 and 18.3% in May 2020, relative to its projected value. The study also investigates system stability by analyzing transmission loss and load factor, and the environmental effect by analyzing the Carbon Dioxide emission rate. Finally, the study provides recommendations for overcoming future disasters, such as developing more resilient power systems, investing in renewable energy, and improving energy efficiency.

Keywords: Bangladesh power system; COVID-19; load forecast; post-COVID.

**List of abbreviations:**

| | | | |
|---|---|---|---|
| COVID-19 | Coronavirus Disease 2019 | PGCB | Power Grid Company of Bangladesh |
| LSTM | Long-Short-Term Memory | IEA | International Energy Agency |
| $CO_2$ | Carbon Dioxide | GWh | Gigawatt hour |
| CEF | Carbon Dioxide Emission Factor | Q1 | First Quarter |
| DSM | Demand Side Management | DG | Distributed Generation |
| EU | European Union | IEDCR | Institute of Epidemiology, Disease Control and Research |
| WHO | World Health Organization | | |

## 1. Introduction

In today's world, the progress of technology is heavily contingent upon the availability of energy, and many recent innovations that attempt to improve people's daily lives would be useless without it. Energy has become an indispensable part of the modern economy, healthcare sector, scientific endeavors, and many more. However, the energy sector is notably vulnerable to natural calamities and outbreaks. These incidents cause interruptions in electricity production and consumption, which in turn affects the reliable operation of the electrical grid. During the Coronavirus Disease 2019 (COVID-19) pandemic, the energy sector of the entire globe encountered its most severe crisis to date. COVID-19, a new virus that emerged in December 2019, stunned the whole world with its rapid spread and high mortality rate. Since its discovery in Wuhan, China, the virus has rapidly disseminated, resulting in the infection and death of millions of individuals, with an estimated 1.25 percent of the world's population contaminated within thirty days [1], [2]. The highest number of deaths had been recorded in the United States, Brazil, India, Mexico, and Russia [3]. The World Health Organization (WHO) declared COVID-19 a worldwide pandemic on March 11, 2020 [4]. Beyond the considerable human casualties, the social and economic consequences were similarly enormous. In response to the pandemic, the majority of nations opted to employ mitigation strategies including the imposition of lockdown measures, travel restrictions, and mandatory closures of businesses and industries. Consequently, a substantial decline in the energy demand was observed. This had a major impact on power grid as large changes in electric demand create challenges for power grid to maintain demand supply balance. Therefore, extensive amount of scholarly literatures had been devoted to examining the impact of the pandemic on electricity sector. Bompard et al. introduced a systematic approach to evaluate the impact of the COVID-19 pandemic on the energy industry of European nations. A power usage decrease of approximately 12.6% was documented for the EU countries. Italy experienced the most substantial declines at -20.9%, followed by France at -18.9%, Spain at -16.9%, and the United Kingdom at -5.2% [5]. Malec et al. analyzed the effect of the pandemic and overall changes in energy usage and demand profiles, based on information obtained from a Polish energy trading and sales company. The data indicated that shopping centers and workplaces experienced the most significant decreases in energy consumption which resulted in a 15-23% decrease in energy usage during the first lockdown in Poland and a maximum of 11% decrease during the second lockdown [6]. Jiang et al. highlighted the changes in energy intensity, stabilizing demand, and focused on the rebound effect of digitalization in energy consumption during COVID-19 [7]. Chattopadhyay et al. investigated the impact of the COVID-19 pandemic on the energy consumption of a residential complex in Hyderabad, India. According to the study; in 2020, residential electricity consumption increased by 15% for the entire day, specifically, it rose by 22-27.50% during the day and 1.90-6.6% at night as a result of COVID-19 [8]. Halbrügge et al. studied the power systems in Germany and Europe during the COVID-19 pandemic, which revealed the net generation of electricity from renewable sources surpassed 55% in the first half of 2020, an increase from the 47% recorded during the corresponding period in 2019. The load demand reduced in all EU countries except Sweden, which did not implement lockdowns [9]. Madurai et al. examined the immediate COVID-19 effects on a few key energy marketplaces around the world and discussed how India's power needs changed between March and May 2020 [10]. Energy demand studies examining the effects of COVID-19 were also documented for the United States [11], China [12], France [13], and Turkey [14]. Most of this literature indicated that a rapid shift in lifestyle due to COVID-19 significantly raised domestic power consumption while decreasing electricity usage in commercial and industrial sectors, ultimately impacting the national energy demand pattern. Despite this, demand patterns vary significantly among different countries. Therefore, it is crucial to conduct region-specific analyses of pandemic-related impacts. Nevertheless, the majority of such studies have predominantly focused on developed countries, thereby neglecting to address the repercussions of COVID-19 on developing countries and the manner in which the energy sector of such developing countries recovered from this disruption.

Bangladesh is a developing nation with a very high population of around 170 million. Due to its large population density (approximately 1180 people per $km^2$), insufficient health care systems, destitution, and sluggish economy, it is considered one of the most susceptible nations to COVID-19. The Institute of Epidemiology, Disease Control and Research (IEDCR) first recognized three instances of pandemic COVID-19 in Bangladesh on March 8, 2020 [15]. Bangladesh recorded 112,306 cases of infection from the entire population, with 1464 fatalities. (till 21 June 2020) [16]. During the initial lockdown period from March 26, 2020, to May 30, 2020, millions of individuals were required to stay in their homes. To contain the spread, the government of Bangladesh had closed all non-essential businesses, offices, schools, institutions, and factories [17]. Consequently, the electricity grid of Bangladesh experienced substantial changes. Bangladesh's electricity sector is among the most energy-intensive industries, with average annual growth in the double digits for the past decade. Therefore, studying the implications of the COVID-19 pandemic on the electricity industry in Bangladesh is critical. However, research in this area is rather scarce. Previously, Alavi et al. [18] analyzed the impact of the lockdown on Bangladesh's total electricity consumption by creating a neural network-based model to forecast energy consumption during the lockdown. Nevertheless, a complete scenario of the impacts of COVID-19 with practical daily, weekly, and yearly data alongside comparison with forecasted data is still missing. Therefore, it is critical to conduct thorough analyses of the energy sector dynamics of Bangladesh during and post-COVID-19 period to evaluate the alterations in demand patterns and to develop preventative measures for the future.

This study investigates the effects of the lockdown implemented to mitigate the transmission of COVID-19 on the electrical power system of Bangladesh. Load demand data are collected from the Power Grid Company of Bangladesh (PGCB) website which is publicly available and these data are used to show how the daily load pattern and yearly load pattern changed and recovered to conventional load pattern during and post-COVID-19 period [19]. The effect on electricity consumption at weekends and weekdays due to COVID-19 is also presented along with the power quality and load factor comparison. A comparison of different fuel-based generations is presented based on COVID-19 and non-COVID-19 periods. A comparison of $CO_2$ emissions for COVID-19 and non-COVID-19 periods is also presented. Again, demand prediction is another sector that has been focused on in this paper. To compare the practical data during COVID-19 with predicted data LSTM-based forecasting is used. The LSTM forecasted load profile excludes the effect of COVID-19 and the difference between actual data and forecasted data is expected to show the severity of COVID-19. Understanding the disparity between actual and projected data is of utmost importance in the context of pandemics such as COVID-19. Additionally, it explores potential strategies to enhance the efficiency of the power and energy sector, as well as grid operations, during similar circumstances in the future. Recommendations are provided to overcome future catastrophes like pandemics or natural calamities (Cyclone).

## 2 Methodology

The methodology for this paper is divided into two parts: the first part consists of data collection and formatting, and the second part consists of predicting the load data of the year 2020 using a machine learning model. Finally, some recommendations are proposed based on the data analysis. The graphical flow diagram of the whole process is presented in **Figure 1**.

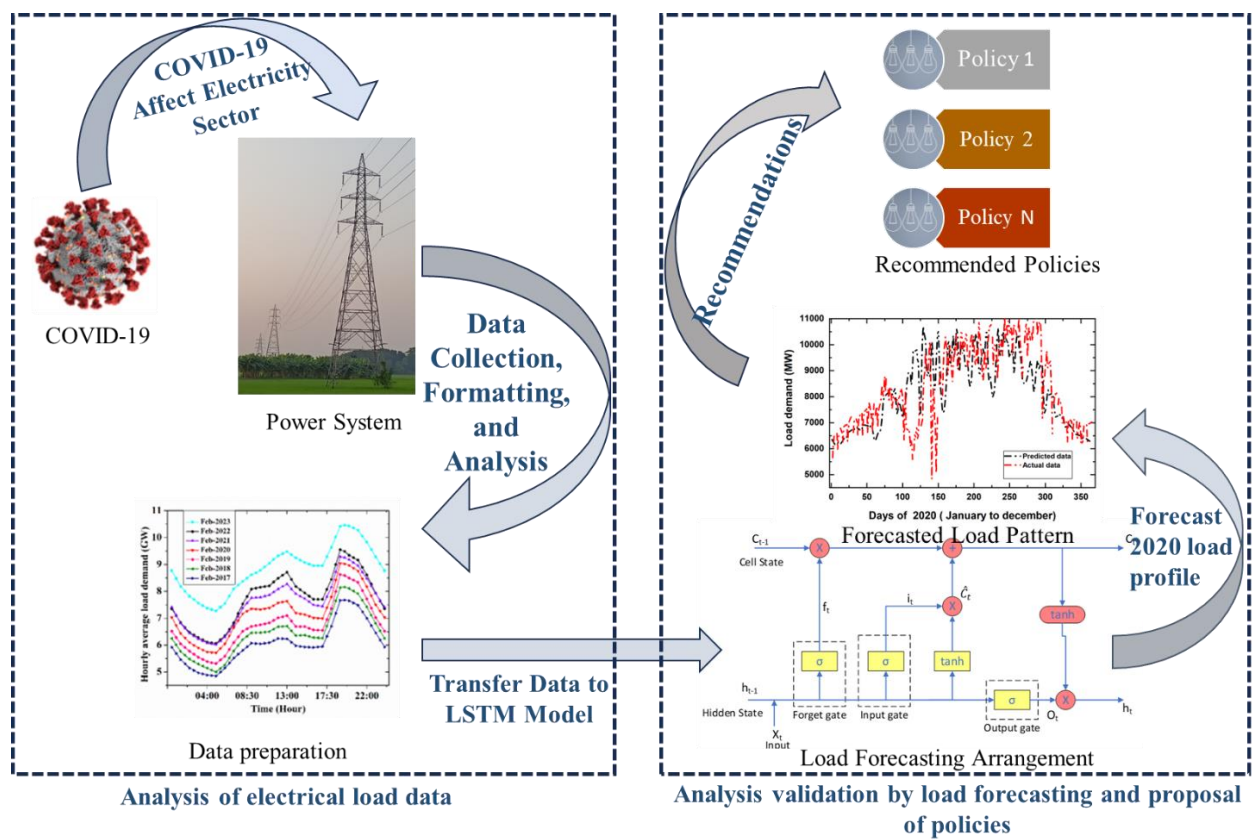

**Figure 1**: Graphical presentation of the flow diagram.

## 2.1 Historical load Data and environmental data collection

As PGCB is the only power transmission entity that exists in Bangladesh, the data on overall daily generation and distribution is collected from the daily transmission report section of the website of PGCB [19]. Data are also available on the website of the Bangladesh Power Development Board (BPDB) [20]. The daily load demand data are formatted in an Excel file for 24 hours and the average load demand/hour is calculated for each particular hour. By considering calculated hourly average load demand data against time for a particular month, the monthly load profile is obtained. To obtain the yearly load pattern the average load demand/hour is calculated and they are placed against the date of that year. As, said in the earlier section the first three cases of COVID-19 were observed in Bangladesh on March 8, 2020 [15], and Bangladesh recorded 112,306 cases of infection with 1464 fatalities till 21 June 2020 [16]. To show the severity of COVID-19, the months of February, March, April, and May are thus considered for different years (2017-2022). The weekend and working days load profiles are calculated by considering the particular days of a particular year from the yearly load profile.

The carbon dioxide ($CO_2$) gas emission from the generation of electrical energy (per kilo-watt hour (KWh) depends upon the nature of the fuel used. The amount of $CO_2$ gas emission in grams/unit for different types of fuel is presented in **Table 1** [21]. From **Table 1**, the average $CO_2$ emission factor (*CEF*) is chosen for a particular fuel, and the total mass of emitted $CO_2$ is found by equation (1).

$$M_{CO_2} = \text{Electrical unit} \times Average\ CEF \tag{1}$$

where, $M_{CO_2}$= mass of $CO_2$, *CEF*= $CO_2$ emmission factor.

**Table 1**: Average $CO_2$ emission factor (gram $CO_2$/unit) for different sources of electricity generation [21] [22].

| Energy generation by | Number of data sources | Minimum *CEF* | Maximum *CEF* | Average *CEF* |
|---|---|---|---|---|
| Gas | 23 | 380 | 1000 | 533.17 |
| Furnace Oil and Diesel | 10 | 530 | 890 | 773.80 |
| Solar | 22 | 13 | 190 | 65.05 |
| Coal | 43 | 660 | 1370 | 942.33 |
| Hydro Power | 12 | 2 | 20 | 8.22 |

## 2.2 **LSTM architecture for load forecasting**

A LSTM-based artificial neural network has been implemented in this work to predict the electricity consumption in Bangladesh for the year 2020. LSTM is a modified version of the recurrent neural network (RNN) that was introduced by Hochreiter et al. [23]. A typical LSTM unit consists of a cell and three gates namely: forget gate, input gate, and output gate as shown in **Figure 2**. The details of the gates with their roles in forming the LSTM network are discussed below:

**Forget Gate ($f_t$):** The formation of the LSTM network starts with the identification of data that are essential to be retained in the memory cell and data that are to be discarded. This is done in the forget gate, $f_t$ which utilizes the current input, $X_t$ at time t, and previous hidden state, $h_{t-1}$ at time t-1. These values are fed in a sigmoid function which returns 0 if the previous information can be discarded by newer, more vital information. On the contrary, 1 is returned when the previous information is preserved. Finally, the sigmoid function result is multiplied by the current cell state, thus eliminating the knowledge that is no longer required as it gets multiplied by zero. The forget gate, $f_t$ thus can be given by the following equation [24]:

$$f_t = \sigma\big(W_f \cdot [h_{t-1}, X_t] + b_f\big) \tag{2}$$

Here, $\sigma$ is the sigmoid function and $W_f$, and $b_f$ represents the weight matrices and bias of the forget gate.

**Input Gate ($i_t$):** After the less important information has been removed the next step involves determining and storing information from $X_t$ in the cell state and updating the cell state. This stage consists of two components: the sigmoid layer and the *tanh* layer. The sigmoid layer makes the decision whether to keep or discard the new information (0 or 1) and the *tanh* function then assigns weights to the input data, determining their level of significance within a range of -1 to 1. These two numbers are finally multiplied to calculate the new cell state. The steps involved in the input gate can be given as follows [24]:

$$i_t = \sigma(W_i \cdot [h_{t-1}, X_t] + b_i) \tag{3}$$

$$\hat{C}_t = \tanh(W_C \cdot [h_{t-1}, X_t] + b_C) \tag{4}$$

$$C_t = i_t \cdot \hat{C}_t + f_t \cdot C_{t-1} \tag{5}$$

Here, $\hat{C}_t$ is the internal cell state, $\sigma$, and *tanh* are the activation functions, $C_t$ and $C_{t-1}$ are the cell state at time t and t-1, respectively. $W_i, b_i$ are the weight matrices and bias of the input gate, respectively.

**Output Gate ($O_t$):** This is the final step where the LSTM model output is determined. The sigmoid function sends the selected information to the output gate, $O_t$ which is then multiplied with the cell state after the cell state is activated by the tanh function. These can be given as follows [24]:

$$O_t = \sigma(W_o \cdot [h_{t-1}, X_t] + b_o) \tag{6}$$

$$h_t = O_t \times \tanh(C_t) \tag{7}$$

Where the weight matrices and bias for the output gates are denoted as $W_o$, and $b_0$, respectively.

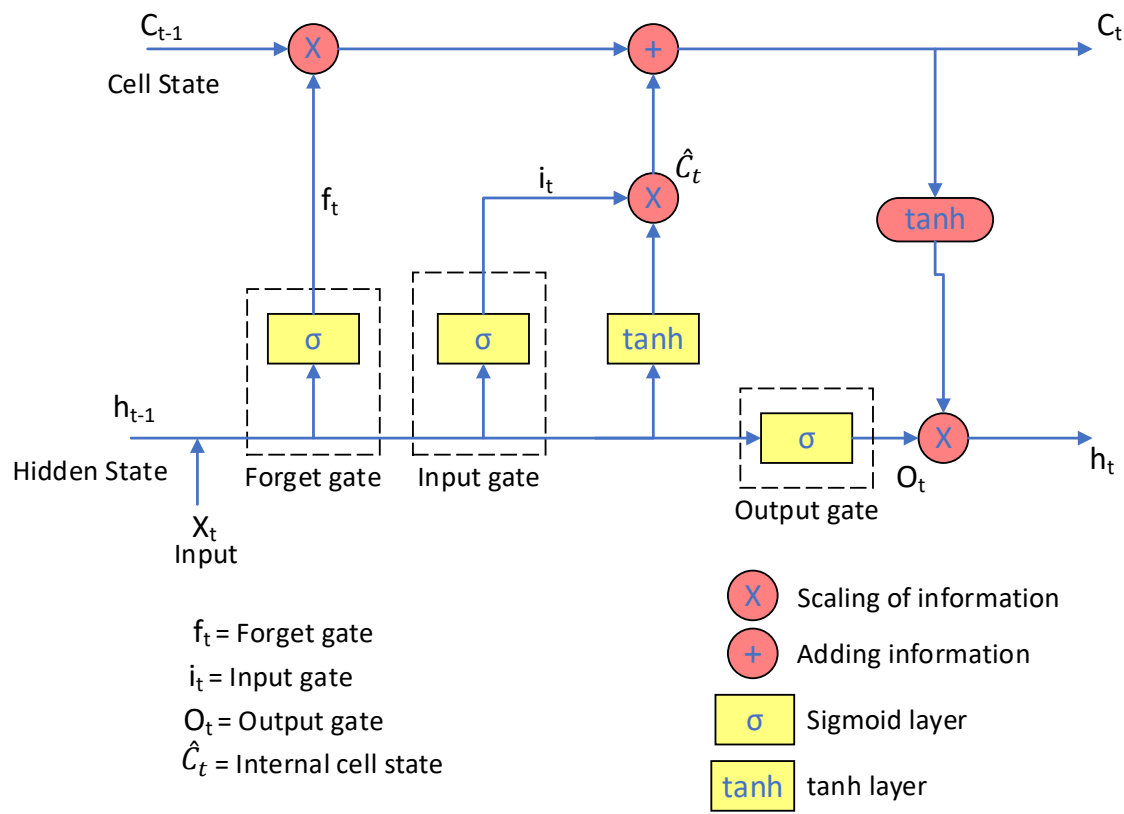

**Figure 2**: LSTM cell architecture

The proposed LSTM architecture in this work consists of four layers of LSTM cells stacked on top of each other. Each LSTM layer has 50 LSTM units. Additionally, dropout layers are inserted after each LSTM layer with a dropout rate of 0.2. Dropout is a regularization technique that helps prevent overfitting by randomly "dropping out" a fraction of the neurons during training [25]. This encourages the model to learn more robust features. Data are preprocessed by adding various time-based features such as month, year, day, and time of day and then resampled to a daily frequency and split into training and test sets. The reason behind the addition of time-based features and then resampling to daily frequency is that it allows the model to capture potential seasonality and periodic patterns in the data. Including these features provides the model with additional contextual information. Daily resampling can help smooth out fluctuations and noise present in the hourly data, making it easier for the model to identify underlying patterns. It also makes the training faster and potentially prevents overfitting to high-frequency noise. The model is compiled with the Adam optimizer [26] having a learning rate=0.001, beta1=0.9, and beta2= 0.999, and the mean squared error (MSE) loss function, which is a common choice for regression problems. Mathematically, the MSE for an LSTM network can be represented as follows [27]:

$$MSE = \frac{1}{N} \sum_{i=1}^{N} (Y_i - Y_{pred,i})^2 \tag{8}$$

where,

$N$ is the number of time steps in the sequence,

$Y_i$ is the true target value at time step $i$,

$Y_{pred,i}$ is the predicted value at time step $i$ produced by the LSTM network.

It's then trained on the training data with a certain number of epochs. During training, the model adjusts its weights and biases using optimization algorithms (e.g., Adam optimizer). The loss function quantifies the model's prediction accuracy, and the optimization process fine-tunes the model to improve its performance. The learning rate is adjusted using a time-based decay schedule, and early stopping is employed to prevent over-fitting. This is how the predicted data is obtained.

The LSTM architecture has been developed in a Python environment by utilizing Tensorflow and Keras deep learning framework. For data preprocessing, Scikit-learn's StandardScalar technique has been utilized. Data manipulation, numerical analysis, and data visualization have been done by utilizing Pandas, Numpy, Matplotlib, and Seaborn libraries, respectively.

## 3. Result and Discussion

There has been a dramatic drop in electricity demand as a consequence of lockdown procedures due to COVID-19 which was discovered in the early period of March 2020 in Bangladesh [28]. Midway through April, the International Energy Agency (IEA) published a study showing that countries in complete lockdown report a 25% drop in weekly energy demands, whereas countries in partial lockdown report an 18% decline [29]. The COVID-19 pandemic has caused a global recession including Bangladesh. During the lockdown, some businesses stayed closed because of the COVID-19 pandemic [30]. So, the power demand has also gone down during that time. As a developing nation, individuals here were less aware of COVID-19 and have returned to their normal electricity use patterns sooner. [31]. As a result, post-COVID-19 recovery in the case of electricity consumption occurred more rapidly than in developed nations. The analysis regarding the impact of COVID-19 is important to construct more resilient grids capable of enduring future interruptions like as pandemics, natural disasters, or economic crises. Also, this analysis informs government policy on the power industry, including encouraging renewable energy, helping vulnerable communities, and grid modernization. This paper analyzes the effect of COVID-19 on the daily load pattern, energy consumption nature of human beings, carbon dioxide emission, yearly load pattern changes, transmission loss, load factor of the overall country, etc.

## 3.1 Effect of COVID-19 and recovery analysis from hourly average daily load pattern

By averaging the load for each hour for a particular month of a particular year against time hourly average daily load pattern is obtained which is depicted in **Figure 3**. The curvature is irregular because the demand is lowest at 5 a.m. as most people of Bangladesh fall asleep at that time and industries also get shut down. After that, the load demand shows an increasing nature and during lunch break again it shows decreasing nature. The working hours again start after the lunch break and load demand increases again. In the evening period, it shows the highest load demand as the lightening household load demand shows its peak value at that time. After that, it goes to decreasing nature again as the industries close and people go to sleep.

The usual rising nature of load demand from February 2019 to 2020 indicates the COVID-19-free period depicted in **Figure 3(a)**. However, the power demand in February 2021 is less than in February 2020 indicating the effect of COVID-19 on the consumption of electricity in Bangladesh. The reason behind this is, February 2020 is free from COVID-19. In early February 2021, Bangladesh, a lower-middle-income country, launched a national vaccination campaign against COVID-19 and within August 14% of people registered for the vaccine were on target [32]. For the vaccination program, people gained immunity against COVID-19. As a result, the power demand of February 2022 is a little bit greater than February 2021 and February 2023 is almost free from the effects of COVID-19. As said earlier COVID-19 hit Bangladesh in the early period of March 2020, and the effect on the electricity sector is more clear in **Figure 3(b)** which shows the load pattern of the year March, 2020 touched the load pattern of March, 2019 at some point and they are very close. It may have a detrimental effect on network hardware, including transmission lines, substations, distribution transformers, and protection devices. To prevent the spread of COVID-19, the government of Bangladesh imposed a lockdown over the country. All shopping complexes and megamalls, grocery shops, educational institutes, government offices, and non-government offices were shut down [33]. Most of the industry had gone through limited production or a complete shutdown [34]. As a result load pattern of April 2020 is recorded to go under the pattern of April 2019 and April 2018. Even at some point, it crossed the April 2017 load pattern which is shown in **Figure 3(c)**. The average power demand fell by 15.4% in April 2020 compared to April 2019. On the other hand, load patterns of April 2022 followed the usual growing nature depicted post-COVID-19 normalization. It is also clear that the April 2020 load pattern is the most irregular curvature in nature. As the lockdown persisted until 30 May 2020, the same scenario is observed in **Figure 3(d)**. There was a significant drop in power demand in May 2020 due to the COVID-19 pandemic which is 17.2% less than in May 2019.

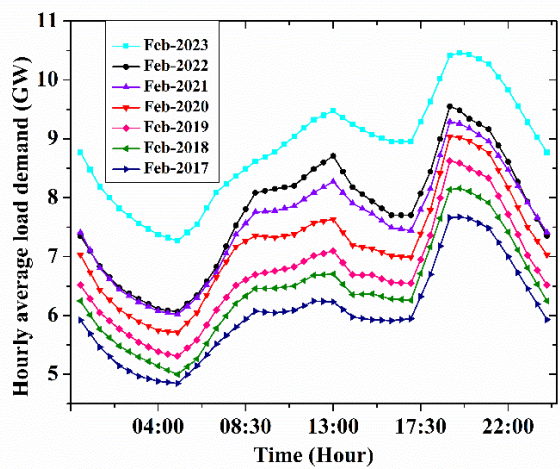

(a)

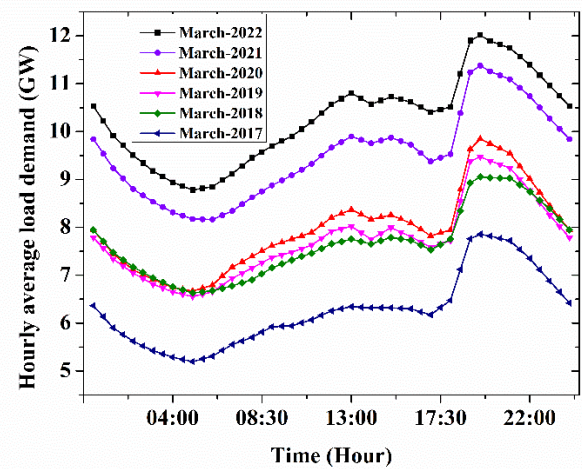

(b)

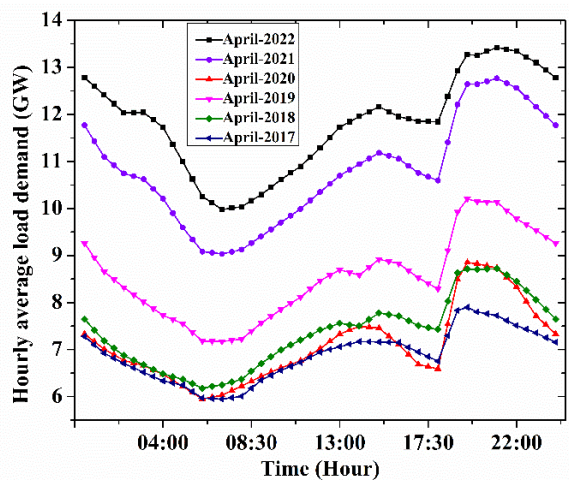

(c)

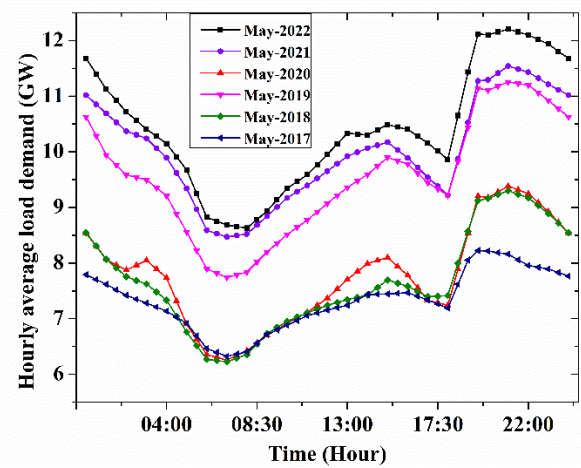

(d)

**Figure 3**: COVID-19 effect and recovery analysis based on hourly average load demand for the months of (a) February  (b)March (c) April (d) May.

### 3.2 Changes in electrical energy generation

The net energy generation of February to mitigate load demand for the years 2019, 2020, 2021, and 2022 are 4048, 4627.87, 4624.42, and 4754.89 Giga Watt-hours (GWh) respectively. The changes in electrical energy generation for a particular month of different years are presented in **Table 2.** The months of February, March, April, and May for the years 2018 and 2019 follow an increasing trend in energy generation which is also observed in 2022. On the other hand in April 2020 energy generation decreased by 1826 GWh due to COVID-19. It can be said, that the year 2022 is fully free from COVID-19 in the case of the electricity sector.

**Table 2:** Changes in electrical energy generation due to COVID-19 [19].

| Month / Year | Changes in electrical energy generation (GWh) | | | |
|---|---|---|---|---|
|  | February | March | April | May |
| 2018 | +573 | +573 | +573 | +504 |

| 2019 | +697 | +697 | +697 | +884 |
|------|------|------|------|------|
| 2020 | +580 | **+234** | **-1826** | **+226** |
| 2021 | **-4** | **+43** | **+3433** | **+43** |
| 2022 | +131 | +476 | +770 | +1246 |

Several European countries and the United States have stated that they plan to reopen parts of the economy in May, April may be the most severely affected month [35] which is seen in the case of Bangladesh also. PGCB transmitted 4666.91 GWh, 5262.56 GWh, and 6123.31 GWh of electricity in April 2020, May 2020, and June 2020, respectively. Compared to the same time last year, the amount of power transmitted is 5491.06 GWh, 6365.85 GWh, and 6082.71 GWh. From April to June 2020, the COVID-19 outbreak caused the total amount of power transfer to drop by 10.51% or 1886.85 GWh. If there hadn't been a COVID-19 spread, for the period in issue, PGCB could have gained Taka 55.14 crore in wheeling revenue [19].

## 3.3 Effect of COVID-19 and recovery analysis from yearly load pattern

The daily average load variation for 2020 is indicated by the red line depicted in **Figure 4**, which follows the same pattern as 2019, 2018, and 2017 until the middle of March before COVID-19 strikes. After that, the government of Bangladesh imposed the lockdown and this effect becomes clear by the red line. Until 09 May 2020, the red line flows under the black line which is an indication of the daily average load flow of 2019. Even at some points, the red line intersects green and blue lines. This happens due to the closure of government and non-government offices, educational institutions, non-essential industries, commercial entities, factories, industries, etc. Beginning on May 10, the government approved the operation of stores and malls on a restricted basis [36]. Also, transportation is limited during that time as a result easy bike (Electric Vehicle) charging is decreased which can affect the load consumption. However, in order to make up for the scholastic loss, online courses for educational schools are launched following a significant study break [37], and electricity demand increases after that time (midway through May). On 18 May, cyclone Amphan made landfall on West Bengal and Bangladesh**.** In Bangladesh, approximately one crore people lack access to power [38]. Electric posts are shattered, wires are shredded by falling trees, and transformers are malfunctioned. Approximately 25000 transmission cables have ruptured. Approximately 300 poles are broken [38]. Due to the cyclone, the country's daily electricity consumption dropped from 10 GW to approximately 2.5 GW. At this time daily average demand falls to 4.8 GW which is depicted in **Figure 4**. The average daily demand starts to increase after 25 May but still, it is under the load curve of 2019. The reason behind this is the government started removing the limitations on all state and private agencies, companies, and factories on May 31 [39]. Beginning on June 1st, public transportation is permitted to operate with 50% occupancy and a 60% fare increase. Domestic flights are also allowed to operate. Movement restrictions are enforced between 10 p.m. and 5 a.m. from July 1 through August 31. On July 12, however, Hafizia Madrasas are permitted to open. On August 8, Islamic higher education institutions are permitted to open. Qawmi Madrasas are eventually permitted to operate on August 24 [40].

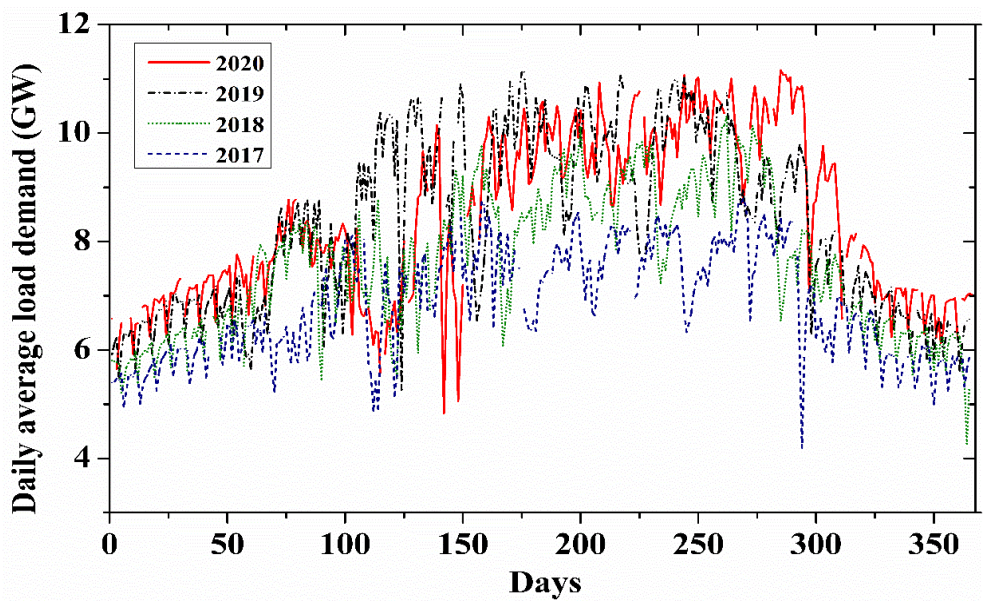

**Figure 4**: Daily average load consumption comparison for the years 2020, 2019, 2018, and 2017.

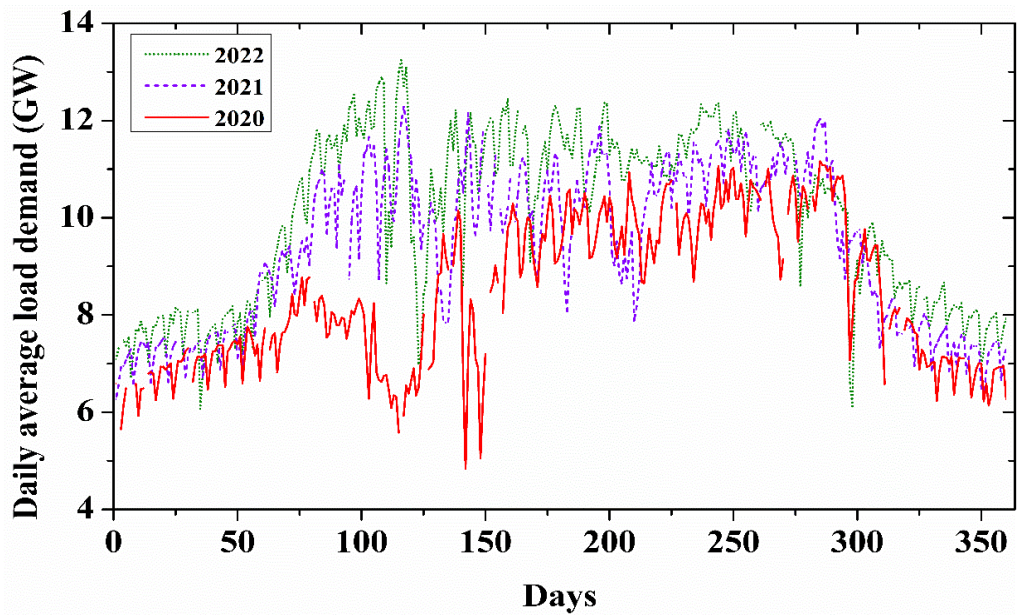

**Figure 5**: Daily average load consumption comparison for the years 2020, 2021, and 2022.

After August 2020 load curve gets back to its as-usual nature. Bangladesh approved the Chinese vaccine trial on August 27th [41]. From December 2020 to the end of February 2021, the infection rate is at its lowest since the outburst of the epidemic. The first vaccinations were administered on January 27, 2021, and the widespread vaccination campaign began on February 7, 2021 [42]. For these reasons, people roll back to there as usual life. Consumptions of electricity nature roll back to their conventional growth nature which is depicted in **Figure 5**. Usually in Bangladesh load demand is higher in summer (April to September) than in winter (October to January) season. This seasonal variation is also observed in **Figures 4** and **5**.

### 3.4 Effect of COVID-19 and recovery analysis from the weekend and weekday load patterns

In **Figure 6**, we examine the effect of COVID-19 on the electricity demand during weekly holidays (Friday, and Saturday) when the bank, schools, and other government nongovernment entity is closed. Therefore people at this time stayed at their homes and the residential load increased. The red line stays at a higher position comparatively than the red line in **Figure 7** which indicates the load profile of 2020 on working days (Sunday to Thursday). Also, the red dots and green triangles are very close to each other showing that the effect of COVID-19 is more significant during weekly holidays from March to November. However, in **Figure 7**, from March to September the effect of COVID-19 is observed. On the other hand, after the September load nature regains its conventional nature as most of the organization start to run again and people are getting relief from the panic caused by COVID-19. This flow continued till now and overall conditions are getting normal which is observed from the black box in both **Figures 6** and **7**.

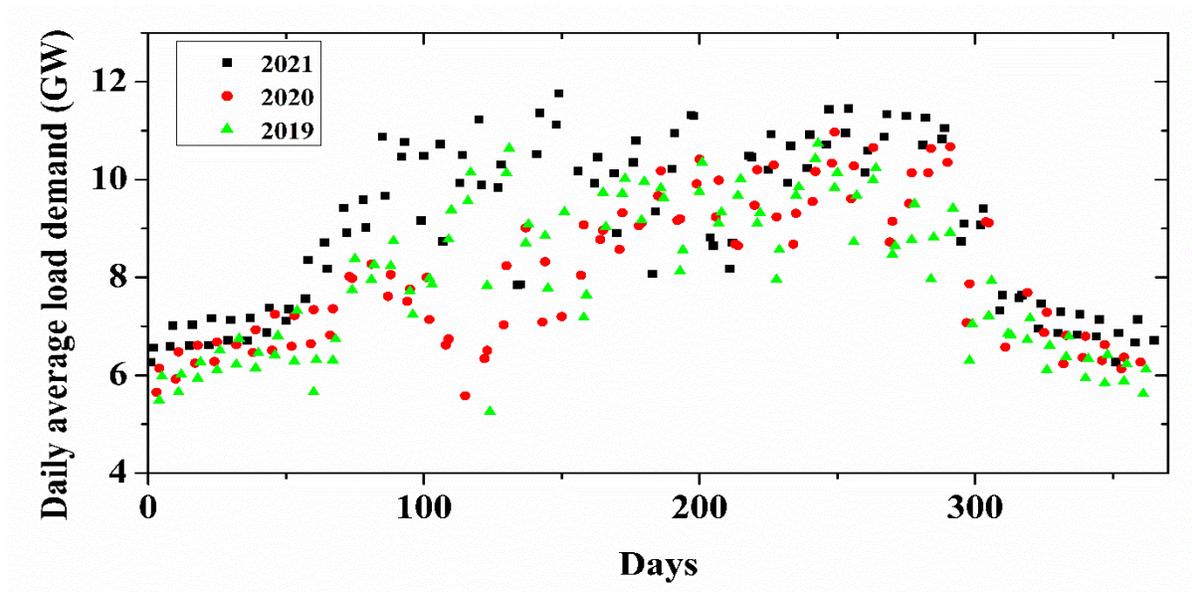

**Figure 6**: Daily average load consumption comparison on weekend days

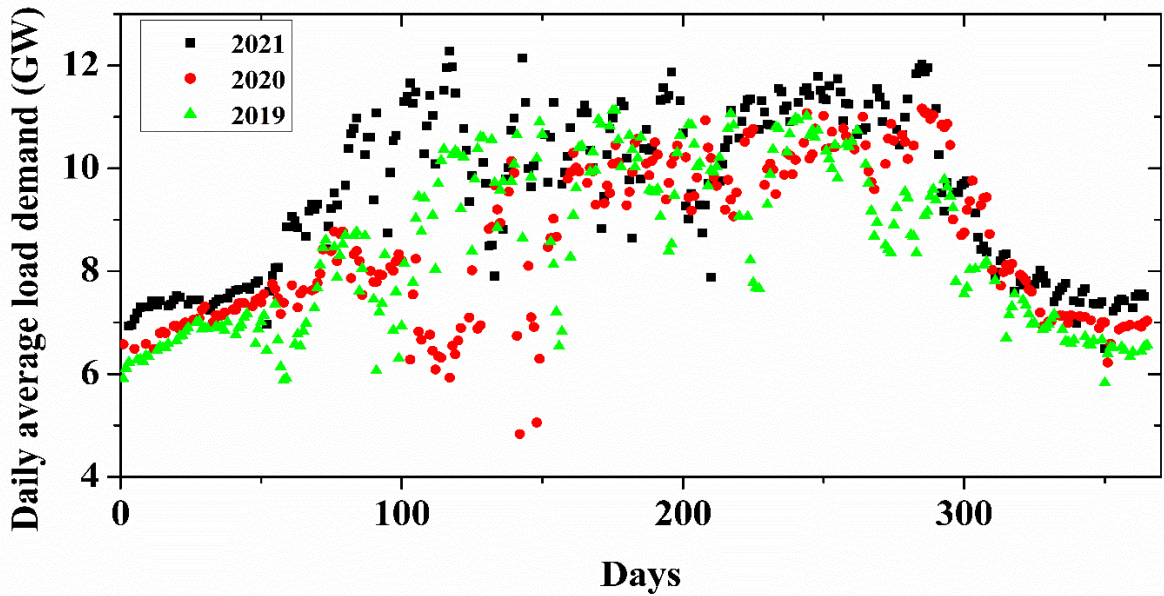

**Figure 7**: Daily average load consumption comparison on working days

## 3.5 Effect of COVID-19 and recovery analysis from changes in transmission loss and load factor

The COVID-19 pandemic has also affected the maintenance and repair of power plants and the construction of new power plants. The pandemic has caused delays in the delivery of equipment and supplies for power plants, which has affected the timely completion of projects. From **Table 3**, The transmission loss has a decreasing nature from the 2015-2016 financial year. However, due to COVID-19, it has shown a growing nature from the financial year 2019-2020 as the transmission system faces stress due to the irregular load pattern. And from the post-COVID-19 phenomenon, again shows a declining nature.

**Table 3**: Comparison of Transmission losses for the different financial years [19].

| Financial year | Transmission loss |
|---|---|
| 2015-2016 | 2.86% |
| 2016-2017 | 2.67% |
| 2017-2018 | 2.60% |
| 2018-2019 | 2.75% |
| 2019-2020 | 2.93% |
| 2020-2021 | 3.05% |
| 2021-2022 | 2.89% |

The load factor is the ratio of average load to maximal demand during a specified period. A large load factor indicates that the electrical load is utilized more effectively. A high load factor results in greater electrical energy saving reduces the peak load demand and, reduces the average cost per unit (kWh). From **Figure 8**, the load factor for the pre-COVID-19 period (January, and February 2020) is nearly 74 and 75 percent. After the introduction of COVID-19, the load factor starts decreasing and has a significant drop in May. 66.46% load factor is the lowest value from the last 8 years to now which is recorded due to COVID-19.

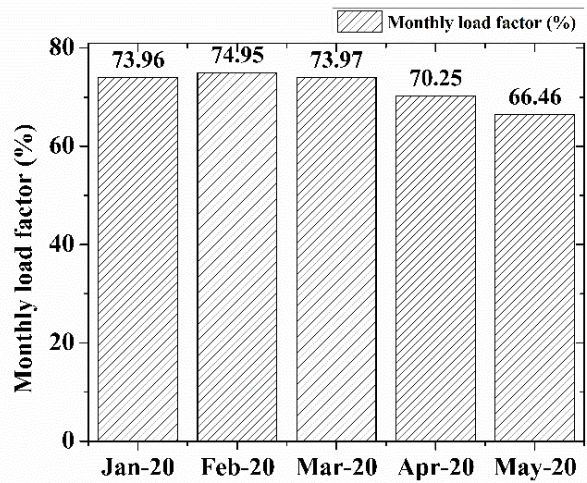

**Figure 8**:  Load factor variation for the first wave of COVID-19 [19]

## 3.6 Effect of COVID-19 and recovery analysis from changes in fuel consumption and $CO_2$ emission

Coal, natural gas, and nuclear power all see a decrease in demand for energy as reported by [29], worldwide $CO_2$ emissions fell by over 5% in the first quarter (Q1) 2020 compared to Q1 2019, with a drop of 8% in coal releases, 4.5% in oil releases, and 2.3% in natural gas releases being the primary causes. The market for coal around the world has largely decreased and it has dropped by almost 8% from the Q1 of 2019.

First-quarter oil consumption is down nearly 5% as well. Gas-based economies are not severely impacted in the first quarter of 2020, so the pandemic's effect on gas consumption is more muted at around 2%. With higher installed capacity and prioritized dispatch, renewables are the only source to report a demand increase [29]. In the case of Bangladesh, from **Figure 9**, natural gas-based generation decreases during the COVID-19 period as a result greenhouse gas emissions are also reduced.

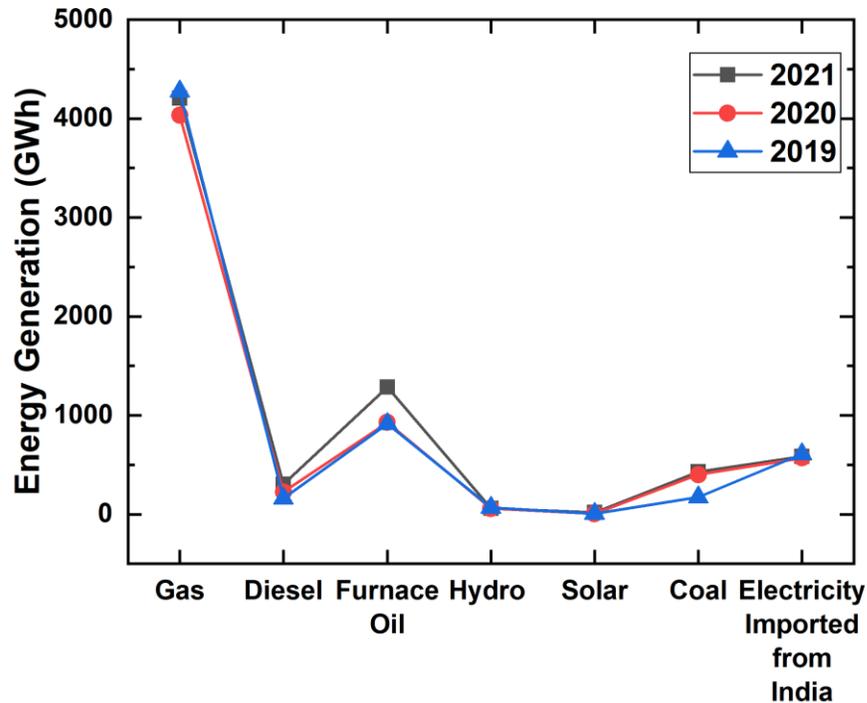

**Figure 9**: Energy generation comparison based on different types of fuel

On the other hand, diesel, furnace oil, and coal show an increasing nature and therefore the emission of greenhouse gases is also in increasing nature. **Table 4** represents a detailed comparison of the year 2019-2021. The power imported from India is reduced during the COVID-19 period. Although the generation of renewable energy increased in developed countries, in Bangladesh, it showed a declining nature. However, in post-COVID-19 duration that belongs to June 2020 from now, it again shows a rising nature. The only hydroelectric power plant also goes through less electricity generation during the COVID-19 period and the generation is reduced by 10.2 GWh compared to the year 2019.

**Table 4**: Comparison of energy generation by different sources [19].

| Energy generation by | 2021 | 2020 | 2019 |
|---|---|---|---|
| Gas (GWh) | 4207.443 | 4033.20833 | 4274.19332 |
| Diesel (GWh) | 303.51026 | 224.1975 | 161.34562 |
| Furnace Oil (GWh) | 1285.10725 | 927.58333 | 914.08995 |
| Hydro (GWh) | 57.14505 | 58.39 | 68.59042 |
| Solar (GWh) | 18.63133 | 6.06917 | 6.31203 |
| Coal (GWh) | 428.6752 | 400.23083 | 172.85628 |
| Electricity Import from India (GWh) | 586.38808 | 570.13167 | 608.20707 |

To generate electricity, $CO_2$ emission has occurred. Although the rate of $CO_2$ emission by gas is reduced during the COVID-19 period, oil and coal show a reverse nature. This rate is expected as large power plants like the first unit of Ultra Supercritical Technology based Pyra power plant came into operation in May 2020 and the second unit in October 2020 [43]. The whole $CO_2$ emission statistics are presented in **Table 5** which is calculated from equation (1) for the year 2019-21. However, the nature of growth in fuel consumption is irregular in 2020, but in 2021 post-COVID-19 comeback is observed through the growing nature in fuel consumption and $CO_2$ emission which is presented in **Table 5**.

**Table 5**: Statistics of $CO_2$ emission by different fuel sources.

| Year<br>Sources | 2021 | 2020 | 2019 |
|---|---|---|---|
| $CO_2$ emitted by Gas (Kilo Ton) | 2243.282 | 2150.385 | 2278.872 |
| $CO_2$ emitted by Furnace Oil and Diesel (Kilo Ton) | 1229.28 | 891.24 | 832.18 |
| $CO_2$ emitted by Solar (Kilo Ton) | 1.212 | 0.395 | 0.411 |
| $CO_2$ emitted by Coal (Kilo Ton) | 403.954 | 377.149 | 162.888 |
| $CO_2$ emitted by Hydro Power | 0.47 | 0.48 | 0.56 |

### 3.7 Effect analysis of COVID-19 from forecasted data

The actual load demands for the year 2020 with predicted data based on long short-term forecasting are depicted in **Figure 10**. For the data prediction, hourly load demands from 2014 to 2019 are provided, and data for 2020 is predicted based on these where assumptions are taken as: the year 2020 is free from the impact of COVID-19 and lockdown. Also, during LSTM data preprocessing, any kind of missing data has been assumed by linear interpolation. The primary objective is to determine the effects of COVID-19 and the lockdown on the load demand of Bangladesh in 2020. From 2014 to 2019, Bangladesh's power sector grew steadily. As these data are used here to predict the load demand in 2020, it is reasonable to assume that the predicted data will continue this load demand growth trend, as the impact of lockdowns is not considered in this prediction. The actual load requirements for 2020, however, are significantly impacted by COVID-19. As shown in **Figure 10**, there is a significant discrepancy between the actual data and the predicted data especially in the range of 80 to 150 days. Which are from the month of March to the month of June 2020. COVID-19 first arrived in Bangladesh in March, and a lockdown was implemented on March 20. Therefore, it is only plausible that this anomaly is primarily caused by COVID-19.

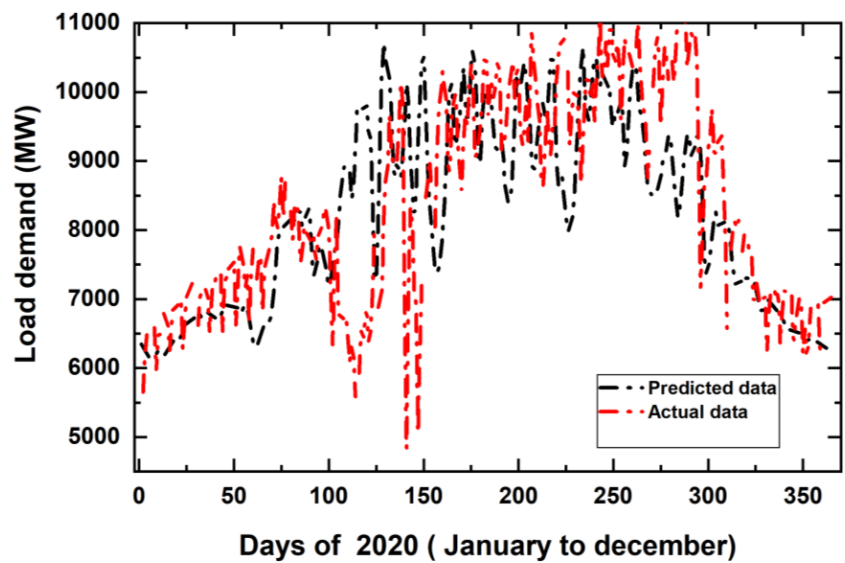

**Figure 10:** Comparison of actual load demand with LSTM based machine learned predicted data of Bangladesh for the year 2020

A breakdown of the load data comparison by month from March 2020 to June 2020 is done to get a clear picture. These data are shown in **Figure 11**. As seen in **Figure 11(a)**, which shows data for March 2020, the actual data initially exceeds the predicted data. However, starting on March 20, the actual data starts to diverge from the anticipated data. While the predicted data follows an upward trend, the actual data for April 2020, shown in **Figure 11(b)**, continues to show a clear downward trend in actual load demand. At this time, the nation is placed under strict lockdowns, and all businesses, colleges, and schools are closed. The actual average load demand had a 19.50% drop compared to the predicted load demand. It appears that the load demand rises a little in May 2020, and starts to decline again at the end of the month because the country is hit by a cyclone called Amphan. This is shown in **Figure 11(c)**. An average decrease of 18.3% in demand is seen for this month. The data comparison for June 2020 is shown in **Figure 11(d)**. At this point, vaccination campaigns are also starting, and there is a significant increase in public awareness of COVID-19. After May 2020, the lockdown is lifted, and businesses and offices start operating in hybrid mode while taking precautions. As a result, after the post-COVID-19 declination, the load demand also increased. The actual average load demand for June 2020 exceeded the predicted average load demand data by 3%. The data from June 2020 make this abundantly clear as the actual data here caught up to the anticipated data. This amply demonstrates the effects of COVID-19 and the lockdown on Bangladesh's load demand and hence, emphasizes the need for implementing various policies to face such a crisis in the future.

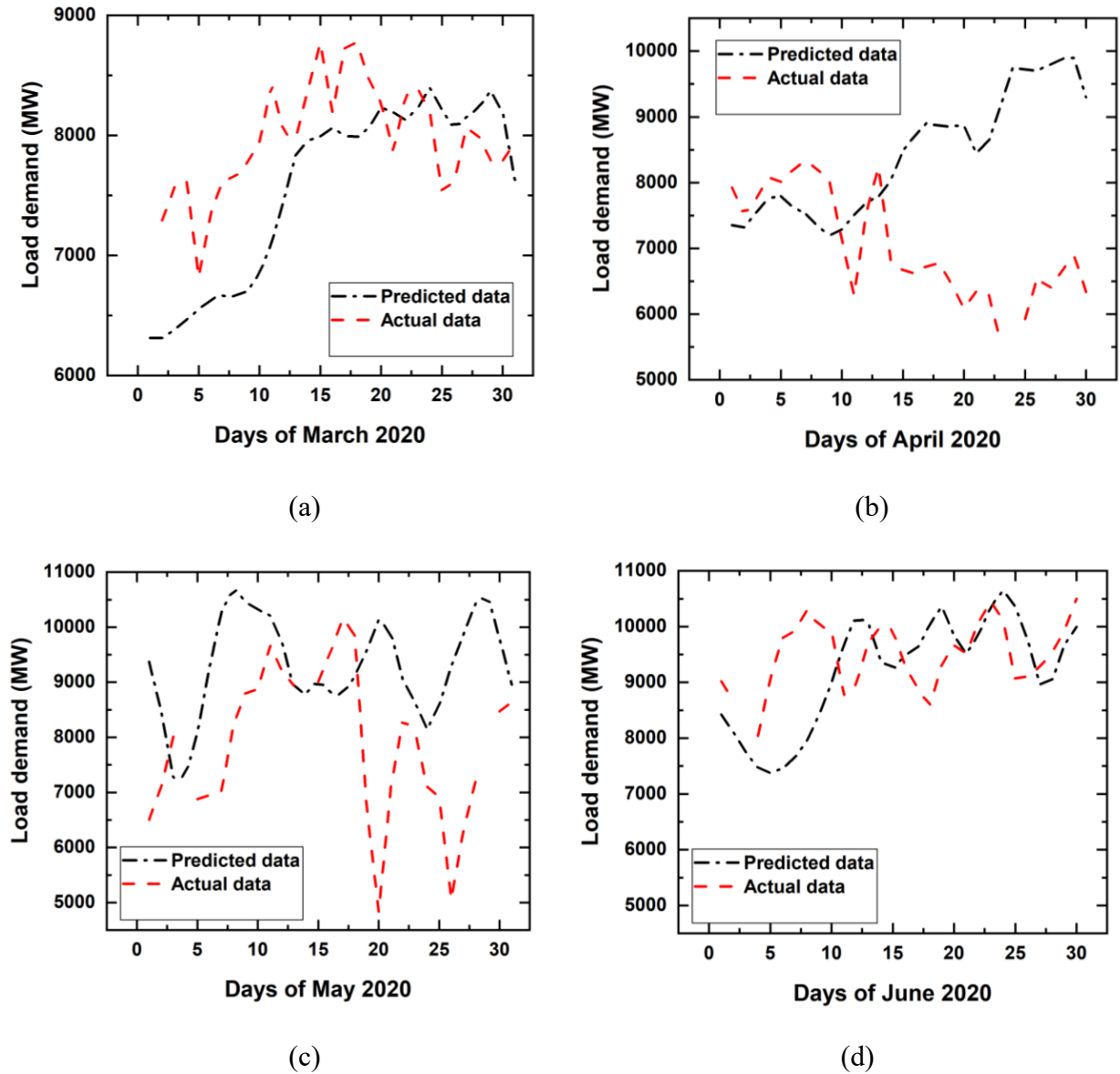

(a)

(b)

(c)

(d)

**Figure 11**: Breakdown of load demand data comparison by months: (a) March 2020, (b) April 2020, (c) May 2020, and (d) June 2020.

## 4. Recommendations of policies for future pandemics

COVID-19 has a significant impact on electricity demand, load profiles, transmission loss, greenhouse gas emissions, and load factors. The following measures can be taken to improve the stability and reliability of the power system for future pandemics:

**Policy-1: Successful implementation of Demand Side Management (DSM):** DSM is a technique that schedules electrical load and flats the load curve. It is seen that during the COVID-19 period business, industry, school, and commercial load decreased due to the lockdown on the other hand, the residential load increased due to the confinement of people in their homes, the increasing use of electronic media, and online activity. The outcome is an irregular load pattern from the previous discussion. In that case, the successful implementation of DSM can be a solution. Proper construction of policy for the implementation of DSM can maintain the load pattern that is best suited for grid stability.

**Policy-2: Laying of underground cable:** Cyclone Amphan traversed Bangladesh on May 18, 2020, and a significant portion went out of power [38]. Approximately 25,000 transmission cables were ruptured and 300 poles were broken. Mainly the overhead transmission line got damaged. So authorities can take policy for laying underground transmission lines as such calamities attack Bangladesh frequently.

**Policy-3: Development of smart grid technology:** During COVID-19 load factor of Bangladesh dropped to 66.46% and transmission loss increased to 3.05%. This type of phenomenon can be occurred in future due to pandemics and other natural calamities, if proper mitigation techniques are not adopted.Policies regarding smart grid technology development can be a good solution for this. Smart grid technology can assist in enhancing efficiency and flexibility in the power system by allowing electrical providers to monitor and regulate the grid in real-time. This can assist in minimizing transmission losses and increase the reliability of the power supply.

**Policy-4: Successful implementation of Distributed Generations (DGs):** Another way to improve load factor and transmission loss is to implement DGs that ensure system stability by proper distribution of generating units throughout a specific region. DGs will help by reducing the need for long transmission lines, supporting voltage levels in areas with high demand, integrating renewable energy sources, and thereby helping to fight future pandemics.

**Policy-5: Improvement of existing infrastructure:** It is seen that the residential load has increased during the lockdown period therefore the power and switchgear equipment related to the distribution side have to be improved to avoid overload.

By taking these measures, it can be expected that a more reliable and resilient power system will form that can withstand future disruptions. The power system needs to be made more flexible and adaptable to changing demand patterns and disruptions. DSM, smart grid technology, DGs, and underground transmission lines can all play a role in achieving this goal.

## 5. Conclusions

This study investigates the impact of COVID-19 on the power sector in a developing country, Bangladesh. The study finds that the pandemic led to a significant decrease in power demand, irregular load profiles, an increase in transmission loss, irregular emission of greenhouse gases, and a low load factor. The government's lockdown measures also have a significant impact on the power sector, as they led to a decrease in commercial, industrial, and school load, and an increase in residential load. The impact of these changes was most pronounced between March 20, 2020, and May 31, 2020, evident in the monthly, yearly, weekly, and weekday load patterns. There is a 1.45 GW power demand decrease in April 2020 and 1.76 GW in May 2020 compared to the same time of the year 2019. In the case of energy demand, there is an 1826 GWh decline in energy demand observed in April 2020 compared to the same month of the previous year. Subsequently, a post-COVID-19 recovery period ensued that began in June 2020, and by 2022, the situation has largely normalized.

Our implementation of a machine-learned load forecast model vividly illustrates the severity of the COVID-19 impact, utilizing load profile data from 2014-2019 to predict the pattern of 2020 without considering COVID-19. The forecasted data verifies the effect of COVID-19 by ensuring a large gap between forecasted data and actual data during the most affected period considering energy consumption (March 2020 to May 2020). The constraints imposed by data availability and accuracy in a developing nation warrant future research to incorporate larger datasets for a clearer distinction between forecast and actual data. Despite these challenges, Bangladesh's authorities demonstrated remarkable acumen in managing the situation, ensuring relatively stable grid operations.

The findings of this study have important implications for developing countries. Developing countries can use these findings to improve the resilience of their power sectors to future shocks, such as pandemics. For example, developing countries can invest in renewable energy, distributed generation, and storage technologies to reduce their reliance on fossil fuels and make their power grids more resilient. Developing countries can also invest in research and development to develop new technologies and solutions to improve the resilience of their power sectors. In the future scope of this paper, researchers can build a model of the DSM and DG implementation strategy showing how these techniques stabilize the grid in case of irregularities. Also, they can show the contribution of smart grid technology in grid stabilization when irregular load patterns strike.

## Declarations

## Author Contributions

A. A.: Conceptualization, methodology, software, validation, formal analysis, investigation, resources, data curation, writing—original draft preparation, writing—review and editing, visualization,. A. A. S.: methodology, software, validation, formal analysis, investigation, resources, writing—original draft preparation, writing—review and editing, visualization, N. K.R. : Conceptualization, investigation, supervision project administration, N. P. B.: formal analysis, investigation, resources, data curation, writing—original draft preparation,, A.N.: Conceptualization, investigation, funding. All authors have read and agreed to the published version of the manuscript.

**Funding statement**

This research did not receive any specific grant from funding agencies in the public, commercial, or not-for-profit sectors.

**Data availability statement**

Data associated with this study is available online. Electricity data is available at https://pgcb.gov.bd/. Also, formatted data is available at the GitHub repository: https://github.com/anis7336/Impact-Analysis-of-COVID-19-in-Bangladesh-Power-Sector-and-Recommendations-Based-on-Machine-Learning.

**Conflicts of interest**

There are no conflicts of interest to declare.

**Acknowledgments**